\pdfoutput=1

\documentclass[10pt,conference]{IEEEtran}

\usepackage[caption=false,font=footnotesize]{subfig}

\usepackage[T1]{fontenc}
\usepackage[utf8]{inputenc}
\usepackage[english]{babel}

\usepackage{url}
\usepackage[]{graphicx}
\usepackage{color}
\usepackage[usenames,dvipsnames,svgnames,table]{xcolor}

\usepackage{tabulary}
\usepackage{multirow}
\usepackage{booktabs} 
\usepackage{ragged2e}

\usepackage{mathtools}
\usepackage{amsfonts}
\usepackage{amsmath}
\usepackage{textgreek}

\usepackage{algorithm}
\usepackage{algpseudocode}

\usepackage[noabbrev]{cleveref}

\usepackage{blindtext}

\usepackage{comment}
\usepackage{flushend}

\newcommand\mycomment[2]{\textcolor{brown}{\scriptsize{{\textcolor{red}{[#1]}} #2}}}

\graphicspath{ {./figures/} }

\title{ Client-Side Routing-Agnostic Gateway Selection for Heterogeneous Wireless Mesh Networks}

\author{
    \IEEEauthorblockN{
    Emmanouil Dimogerontakis\IEEEauthorrefmark{1}\IEEEauthorrefmark{2}, Jo\~{a}o Neto\IEEEauthorrefmark{1}, Roc Meseguer\IEEEauthorrefmark{1}, Leandro Navarro\IEEEauthorrefmark{1} and Lu\'{i}s Veiga\IEEEauthorrefmark{2}}
    \IEEEauthorblockA{
    \IEEEauthorrefmark{1}
    Universitat Politècnica de Catalunya,
    Barcelona, Spain \\
    \url{{edimoger,jlneto,meseguer,leandro}@ac.upc.edu}\\
    }
    \IEEEauthorblockA{
    \IEEEauthorrefmark{2}Tecnico Lisboa/INESC-ID Lisboa,
    Lisboa, Portugal\\
    \url{luis.veiga@inesc-id.pt}\\
    }
}

\date{August 2016}

\begin{document}

\maketitle

\begin{abstract}
Citizens develop Wireless Mesh Networks (WMN) in many areas as an alternative or their only way for local interconnection and access to the Internet. This access is often achieved through the use of several shared web proxy gateways.
These network infrastructures consist of heterogeneous technologies and combine diverse routing protocols. Network-aware state-of-art proxy selection schemes for WMNs do not work in this heterogeneous environment.
We developed a client-side gateway selection mechanism that optimizes the client-gateway selection, agnostic to underlying infrastructure and protocols, requiring no modification of proxies nor the underlying network. The choice is sensitive to network congestion and proxy load, without requiring a minimum number of participating nodes.
Extended Vivaldi network coordinates are used to estimate client-proxy network performance.
The load of each proxy is estimated passively by collecting the Time-to-First-Byte of HTTP requests, and shared across clients.
%
Our proposal was evaluated experimentally with clients and proxies deployed in guifi.net, the largest community wireless network in the world.
%
Our selection mechanism avoids proxies with heavy load and slow internal network paths, with overhead linear to the number of clients and proxies.
\end{abstract}

\section{Introduction}

The majority of the world's population does not have adequate, if at all, access to the Internet~\cite{Isoc15}. This implies that the Internet cannot provide service to the general public, reaching anyone without discrimination.
Global access to the Internet requires a dramatic reduction in the Internet access costs, especially in geographic areas and populations with low penetration~\cite{Gaia2016}. Wireless Mesh Networks (WMNs)~\cite{Vega2015} allow local communities to build their own network infrastructures, known as Community Networks (CNs), providing affordable inter-networking with the Internet and including isolated rural communities worldwide~\cite{ReyMoreno2013}. Sharing resources, such as infrastructure or Internet access, is encouraged at all levels \cite{ITU2009,EPC2014} to lower the cost of network infrastructures and services.\looseness=-1

Among many others, the guifi.net community network exemplifies how communities can develop their own network infrastructures as a commons~\cite{Baig2015}, using several interconnected WMNs. This results in a large-scale network with heterogeneous network performance, which uses diverse routing protocols in different network zones. 
Moreover, guifi.net exemplifies how participants are sharing several Internet gateways among them for web access, the most popular traffic in CNs. This resource sharing is typically implemented using web proxies, and at a smaller fraction, by IP tunnels.
Guifi.net's registered users (12,500) can use any of the 356 web proxies (May 2016). 
In addition, the network links between nodes are contributed and managed by the participants. Therefore, paths between nodes, such as client to proxy may not be reliable \cite{Akyildiz2005} or guaranteed, especially when compared to commercial offerings from centrally managed ISPs.
Access to the Internet through web proxy gateways relies on users or organizations sharing the full or spare capacity of their Internet connection with other guifi.net users.\looseness=-1

As a consequence of the lack of regulation, and despite being a critical service for the community, current proxy gateway services are quite fragile, especially considering large-scale usage~\cite{Dim17a}. Some proxies may be overloaded and, therefore, offer degraded or unusable performance, while others may remain underutilized. Users of overloaded proxies, or those using congested links to reach their proxy, experience degraded quality of experience (QoE) in web access. It is interesting to note that the set of overloaded and underutilized proxies varies according to the access patterns of the users.\looseness=-1

In this paper we focus on the challenges to improve web access experience in an heterogeneous large-scale inter-WMN community, using a pool of shared web proxies. The challenge is that client-nodes could select the right proxy according to the network path performance and the status of available web proxies. This is related with the net effect, which explores the problem of a large population of C clients who can browse the web taking advantage of the aggregated capacity of a pool of P web proxies, with C $\gg$ P, over a heterogeneous WMN infrastructure, at a fraction of the cost of C Internet connections.\looseness=-1

To tackle this issue, we aim to improve the proxy system, without making any changes that could be incompatible with the existing environment. The solution should be:
\textit{i) Incremental and backwards-compatible} since we should be able to deploy it incrementally, so that it works well for both baseline or enhanced clients.
\textit{ii) Dynamic} since users should switch proxies wisely to maximize their QoE. This can be due to changes in network topology, path load, or proxy performance.
\textit{iii) Decentralized} therefore not requiring any central component.
\textit{iv) Routing-agnostic} thus independent of the transport and routing algorithms, or any specific network features.\looseness=-1

The contributions of this paper are threefold. 
First, we propose and evaluate a mechanism where clients use two latency-based metrics to rank proxies and 
select the top ones in terms of QoE, or to switch to the next best proxy when performance degrades (see Sec.~\ref{sec:evaluation}). 
Second, we evaluate a network performance metric based on the usage of the Vivaldi network coordinates and for external nodes to the Vivaldi network (web proxies), in a heterogeneous wireless network environment (see Sec.~\ref{sec:network}). 
Third, we design and evaluate a web proxy performance estimator based on Time-To-First-Byte (TTFB), which is typically used to measure destination web servers (see Sec.~\ref{sec:proxy}).
The proposed mechanism is client-side, as described in the overview of our approach (see Sec.~\ref{sec:overview}). Our evaluation confirms that our mechanism can avoid hotspots, while maintaining a low overhead as proved in Sec.~\ref{sec:overhead}.\looseness=-1

The metrics and the client selection mechanism were instantiated in guifi.net using the Community-Lab.net \cite{Navarro2016} infrastructure. In this mechanism, nodes are acting as clients interacting with a set of guifi.net web proxies. Experimental results show that our proposal is reliable and effective: our method is able to provide good measures of client-proxy and proxy-Internet latencies, following its variability. We found out that our client selection mechanism is cost-effective in finding proxies that result in good web performance and QoE for users. Our results show improvements in the cost-benefit of our proposal in comparison with other quick-to-measure alternatives (such as Vivaldi-only and minimum hops). Our mechanism also proved to be less costly in traffic and delay than slower performance-oriented measures.\looseness=-1

\section{Related Work}
\label{sec:related}

Our proxy selection problem is strongly related with the topic of gateway selection in wireless mesh networks which has been extensively studied in the past. 
The works presented in \cite{Boushaba2011,Ancillotti2010} fail to function in heterogeneous environments since they present solutions that operate in the mesh routing layer as they require modifications in the infrastructure routers, inherently prohibitive in heterogeneous environments. \cite{Ashraf2009, Galvez2008} require additional software in the side of the gateways.  All the works mentioned, despite proposing interesting solutions, they lack practical implementation or testing in a real environment. An exception to the above, and closer to our work is \cite{Ko2013}, where the clients cooperate to probe the gateways and then use the results to select a proxy. Furthermore, while conceptually \cite{Ko2013} can function in heterogeneous environments, in practice it needs modification of the existing underlying routing protocols. \looseness=-1

Concerning heterogeneous wireless mesh network performance measurements, the majority of the solutions for wireless mesh networks are based on active monitoring of network metrics, such as path delay in \cite{Ko2013,Bortnikov2007}, estimated link quality in  \cite{Boushaba2011,Ashraf2009}, link interference in \cite{Boushaba2011,Ashraf2009} and path packet loss rate in \cite{Ko2013}. All these approaches would entail a high monitoring overhead, except \cite{Ko2013}, where monitoring is done cooperatively to reduce the overhead.\looseness=-1

As far as Internet gateway performance measurement is concerned all the above proposals use active measurements to evaluate its performance. More specifically \cite{Bortnikov2007} uses a congestion delay function, \cite{Ancillotti2010} monitors the unused Internet Connection (available capacity). \cite{Boushaba2011,Ashraf2009} require the gateways to participate in the monitoring process by measuring the queue length of their Internet interface, while \cite{Ko2013} is performing active probes. Contrary to these approaches, our solution is totally passive, implying though less accuracy in exchange of saving scarce network resources.\looseness=-1

%
%
As presented later, we used the Vivaldi \cite{Dabek2004} network coordinates system for estimating network performance. 
From an abstract perspective, network coordinates are a virtual positioning system where nodes gather information about the network to position themselves and other nodes in a coordinate space and are used to estimate the inter-node latency. Vivaldi \cite{Dabek2004} is a fully distributed network coordinates system that functions based on the idea of placing nodes in a two-dimensional euclidean space. The measured \textit{ping} latency between the nodes is used to position them in the euclidean space. In addition to the probing, Vivaldi also uses spring-relaxation to \textit{nudge} nodes in the Euclidean space to minimize prediction errors. While there have been proposals for updates of the Vivaldi algorithm the original algorithm is performing fine compared to the improvements \cite{Chen2011}. Moreover, the state of the art of network coordinates includes more sophisticated and more accurate systems, which nevertheless are not fully distributed since they are based on the idea of the external landmarks, like Pharos \cite{Chen2009}.\looseness=-1 

As it can be seen, while the performance measurement and gateway selection in WMNs are had been studied in the past, the existing approaches cannot be applied in a large-scale heterogeneous environment, as in the presented scenario. 


\section{Overview}
\label{sec:overview}

Our goal is to design a practical, non-optimal but best-effort, scheme where clients (user nodes) can select a proxy using network and proxy performance metrics that would not require the modification of any network components and that could function in a heterogeneous environment. 
To this end, we implemented an estimation-based monitoring framework for proxy selection, where clients cooperate sharing their network and proxy performance estimations in order to prioritize their list of known proxies. This allows clients to be able to make an informed proxy selection decision. 
Unlike other proposals, the framework does not try to find an optimal client-proxy assignment, but helps clients avoid bad choices (overloaded proxies, slow Internet connections or slow internal network path) that would degrade significantly their service experience. The non-optimality is the price we have to pay in order to achieve a scalable and practical solution that can be applied in real heterogeneous WMNs while retaining a low overhead. 
The proposed framework is user-friendly. This means that the users do not manage the proxy selection nor switching. They just need to install our component in their client nodes.
It is also important to note that we do not cover the orthogonal problem of proxy discovery in this work. We assume the set of proxies are known beforehand by the clients.
\looseness=-1

The network performance estimator provides estimates of the client-client and client-proxy network latency. It is based on Vivaldi network coordinates \cite{Dabek2004} and extended in a similar way to~\cite{Ledlie2008} in order to estimate the round-trip latency of nodes that are not part of the Vivaldi network -- the proxies. 
All the clients of the proxy selection system participate in the Vivaldi network, exchanging a small amount of messages periodically, which allows them to maintain an updated view of the latencies across them. 
Moreover, each client periodically has to monitor one of the proxies and share this information with the rest of the clients. As we demonstrate in \Cref{sec:network}, these measurements suffice to allow the clients to create a preference list, which orders the proxies according to their network latency.\looseness=-1

The proxy performance estimator provides estimates of the load of the proxy, concerning the quality of the service currently being provided. It is based on the widely used practical assumption that the TTFB
of an HTTP request can reflect the service performance~\cite{Sundaresan2013,Chen2015}. 
In our framework, each client passively calculates the TTFB of the HTTP replies that he receives from his proxy.
Then the client can use this value to estimate the load of his proxy and share it with his Vivaldi neighbors. 
Notice that cache does not help in this case. There are many proxies with little or no space and most of the requests are not cacheable, as we show in \cite{Dim17a}. In addition, the mechanism is applicable to non-transparent gateways such as NATs. Our proxy load estimator, based on latency (TTFB), is at application layer, so it could perform on any type of gateway that accept HTTP requests.
As we present in \Cref{sec:proxy}, this mechanism allows clients to avoid proxies with heavy load or high delay Internet connections. \looseness=-1

\textbf{System Model}
For the model description we assume a static wireless network topology.
We make no assumptions about the quality of the network, and we allow dynamic link conditions (a very slow link is indistinguishable from a very congested link). Latency is our metric of load, for both links and proxies.\looseness=-1

Let $C$ denote the set of clients (user nodes), and $P$ denote the set of proxies.
For every request that a client $c\in C$ is sending to a proxy $p \in P$, \cref{eq:1} shows the experienced latency model.
\begin{gather}
t_{lat} \approx t_{request\_c\_p} + t_{proxy\_p} + t_{response\_c\_p}\label{eq:1}\\
t_{request\_c\_p} \approx A*t_{mesh\_rtt\_c\_p}\label{eq:2}\\
t_{lat} \approx 2*t_{mesh\_rtt\_c\_p} + t_{proxy\_p} + t_{response\_c\_p}\label{eq:4}
\end{gather}


Where $t_{request\_c\_p}$ represents the time required by client $c$ to connect to proxy $p$ and send the request. It is proportional to the round-trip time between $c$ and $p$, $t_{mesh\_rtt\_c\_p}$.
The $t_{mesh\_rtt\_c\_p}$ latency, \cref{eq:2}, depends on the network conditions of the chosen path between client $c$ and proxy $p$. For the rest of this paper we assume that A equals to 2, which corresponds to the client-proxy TCP handshake and the HTTP request. The $t_{proxy\_p}$ latency represents the total time that proxy $p$ needs to process the request until he initiates the request to the remote server. This includes the time that the request is waiting before starting to be served, which is a good indicator of the load of proxy $p$, as it correlates directly with the number of outstanding proxy requests yet to be served. We assume that at a given point in time different clients experience the same $t_{proxy\_p}$ if they use proxy $p$, independently of who is measuring it. In \Cref{sec:proxy} we validated this assumption. Finally, $t_{response\_c\_p}$ is the time that proxy $p$ takes to complete the HTTP request. This time depends on the load and capacity of the proxy's Internet connection and on the latency to access and retrieve the content, related to the distance from the content and content availability. 
From all the above, we deduce that the request latency can be approximated by \cref{eq:4}.\looseness=-1


We argue that $t_{mesh\_rtt}$ and $t_{proxy}$ can provide clients with a good preference indicator allowing them to avoid loaded proxies and proxies located in slow paths. \Cref{sec:network} describes how we use Vivaldi to estimate $t_{mesh\_rtt}$, while \Cref{sec:proxy} elaborates on how TTFB can be used to estimate $t_{proxy}$.

\textbf{Experimental Environment}
In order to assess our decisions, we experimented separately with each component of our solution. Following the practical approach of our work, we decided to perform our experiments in guifi.net, under real heterogeneous wireless network conditions. For the experiments, we were given access to 5 end-nodes across different guifi.net networks and 3 proxies that are also being used by guifi.net users. The nodes and the proxies are distributed in various locations of Catalonia, Spain. Despite the small scale of our experiments, we are still able to assess the behavior of the presented components. \looseness=-1

As explained in \Cref{sec:network}, proxies do not actively participate in the measurements, they nevertheless, need to respond to UDP pings, allowing clients to estimate their round-trip latency. While for the results presented here we used a UDP echo server in the proxies, obstacle which can be practically overcome with tools such as Scriptroute~\cite{Spring2003}.\looseness=-1 

\section{Network Performance Estimation}

\label{sec:network}

In this section we describe and demonstrate how an extended version of Vivaldi~\cite{Dabek2004} can be used to estimate the current performance of the network, expressed as a latency metric, helping the clients to avoid overloaded paths.\looseness=-1

\textbf{Estimating Latency with Network Coordinates}
Each client in our system participates in a Vivaldi network coordinates system to estimate his  round-trip latencies to the other clients. Based on the clients' network coordinates, we implement the ideas described in \cite{Ledlie2008}, modifying them to provide more accurate estimates and to estimate latencies for external nodes to the Vivaldi network.
We show that this allows the clients to maintain an updated estimation about their latency towards each proxy, but excluding proxies from the Vivaldi network. 
Although Vivaldi was designed to predict latency between hosts in the Internet (mostly wired), we show that it can also be used to predict latencies in WMNs despite the RTT variations caused by the wireless environment.\looseness=-1

Vivaldi estimates RTT by sending ping between nodes. Each Vivaldi node maintains a list of $C+R$ \textit{neighbors}: $C$ that are estimated to be the closest nodes, and $R$ other random nodes, located anywhere in the network. The algorithm works in \textit{rounds}, which are triggered every $T$ seconds. 
In every round, a node randomly selects a neighbor from the list, sends $N$ UDP pings to him, and asks him to send back its own pings and its neighbors' coordinates. The variables $C$ and $R$ can be tuned depending on the size of the network and the topology in order to increase random/remote node discovery or to create strong local clusters. The variable $N$ affects the accuracy of the prediction in exchange for the ping traffic overhead.\looseness=-1

Using the client-client estimations we can satisfactorily predict round-trip latency from a Vivaldi node to each proxy, leaving the proxies unmodified since they do not actively participate in the network coordinates system. In this \textit{extended Vivaldi} version, each Vivaldi node maintains coordinates that represent $C+R$ proxies, as described above. In every round, a node performs $N$ UDP pings to a proxy $p$, which is selected in a similar manner than how the node selects its neighbors. Then, the node updates the coordinates it maintains for $p$ and shares the measured latency with its selected neighbor for this round. Then, the neighbor updates the coordinates that it maintains for proxy $p$ as described in \cite{Ledlie2008}. \looseness=-1

\begin{figure*}[t]
\centering
\begin{minipage}[t]{0.30\linewidth}
\centering
    \includegraphics[width=1\linewidth]{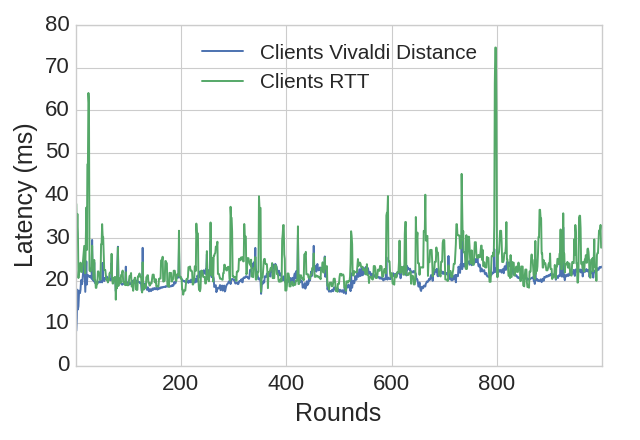}
    \caption{Clients' estimated latency can track RTT behaviour but with various faulty spikes.}
    \label{fig:vivaldi_vs_rtt_rounds}
    \hrule height 0pt
\end{minipage}
\hspace{\columnsep}
\begin{minipage}[t]{0.30\linewidth}
\centering
    \includegraphics[width=1\linewidth]{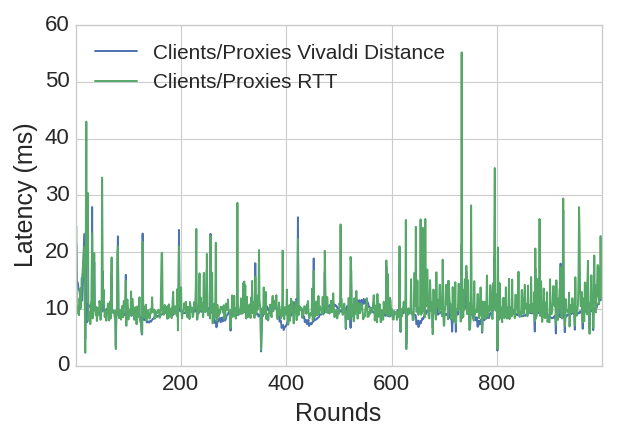}
    \caption{Clients/Proxies estimated latency successfully tracks RTT behaviour.}
    \label{fig:vivaldi_vs_rtt_rounds_proxies}
    \hrule height 0pt
\end{minipage}
\hspace{\columnsep}
\begin{minipage}[t]{0.30\linewidth}
    \centering
    \includegraphics[width=1\linewidth]{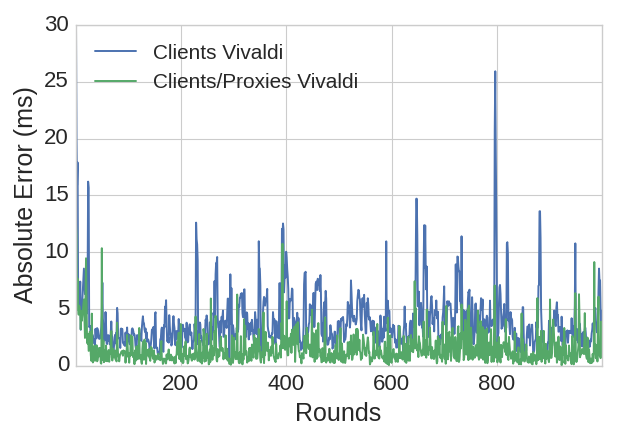}
    \caption{Clients/Proxies estimated RTT error is lower than Clients' RTT error.}
    \label{fig:vivaldi_abs_error_rounds}
    \hrule height 0pt
\end{minipage}
\end{figure*}
\begin{figure*}[t]
\centering
\begin{minipage}[t]{0.30\linewidth}
    \centering
    \includegraphics[width=0.95\linewidth]{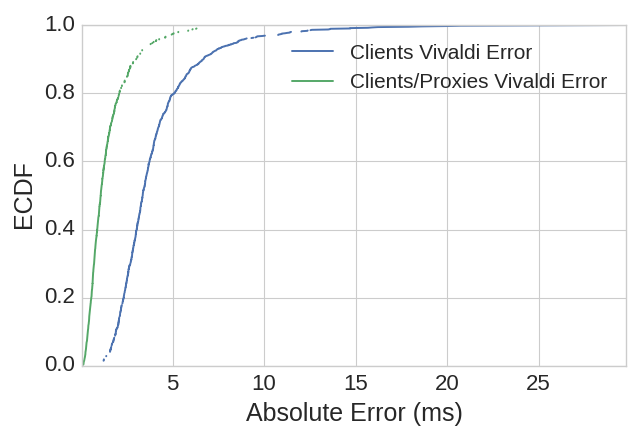}
    \caption{Clients/Proxies estimated RTT error is asymptotically lower than Clients' RTT error.}
    \label{fig:vivaldi_abs_error_ecdf}
    \hrule height 0pt
\end{minipage}
\hspace{\columnsep}
\begin{minipage}[t]{0.30\linewidth}
    \centering
    \includegraphics[width=0.95\linewidth]{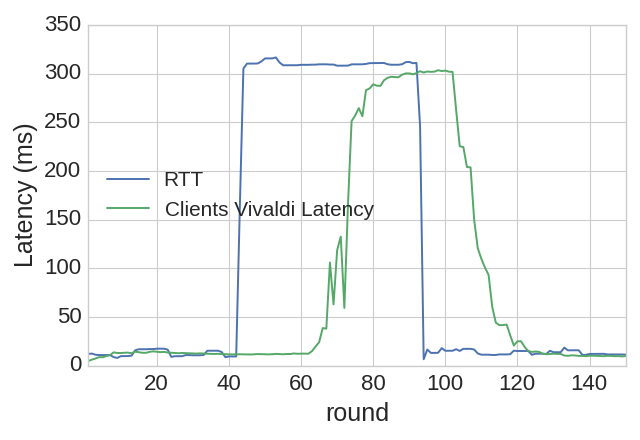}%
    \caption{Predicted Clients' RTT reflects the changes in real RTT for clients but slowly}
    \label{fig:vivaldi_delay_client}
    \hrule height 0pt
\end{minipage}
\hspace{\columnsep}
\begin{minipage}[t]{0.30\linewidth}
    \centering
    \includegraphics[width=0.95\linewidth]{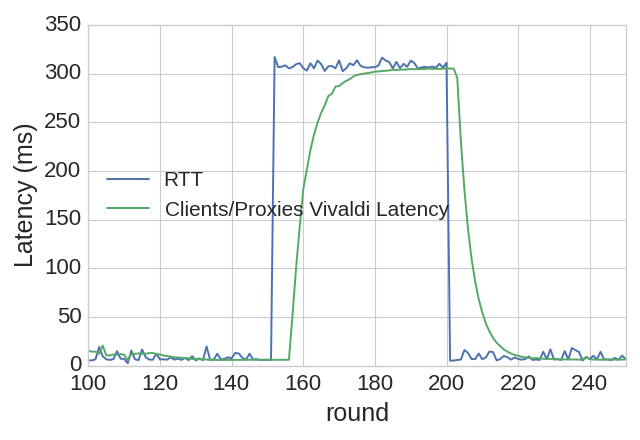}%
    \caption{Predicted Clients/Proxies RTT adapts fast to changes in real RTT}
    \label{fig:vivaldi_delay_proxy}
    \hrule height 0pt
\end{minipage}
\end{figure*}

\textbf{Latency Estimation Evaluation}
For our evaluation, similarly to \cite{Dabek2004}, we define the error of a path as the absolute difference between the predicted RTT for the path (using the coordinates for the two nodes located at the ends of the link) and the actual RTT. 
The error of a node is defined as the median of the path errors for paths involving that node. The error of the system is defined as the median of the node errors for all nodes.\looseness=-1

We experimented, using the described environment, in order to characterize the behavior of the Vivaldi coordinates in an heterogeneous large-scale wireless mesh network. First, we performed an experiment where clients are using Vivaldi to estimate the latencies between them and the extended version of Vivaldi to estimate their RTT to the proxies. This way we can understand the predictive potential of the selected algorithms. It is worth reminding that our experiment was executed in nodes that participate in a real network and therefore were processing real network traffic and using shared links.
\Cref{fig:vivaldi_vs_rtt_rounds,fig:vivaldi_vs_rtt_rounds_proxies} show the real and predicted latencies between clients throughout the experiment.
The median latency between the clients was 22.29~ms, while the predicted median  was 20.82~ms. The median latency between clients and proxies was 9.8~ms while the corresponding predicted median  was 9.36~ms. 
\Cref{fig:vivaldi_abs_error_rounds} depicts the absolute prediction error of the Vivaldi estimation between clients as well as the one between clients and proxies. We observe that the error of the latency prediction between clients and proxies is lower
This can be attributed to the smaller variation of the real latency between clients and proxies, but also to our improvements related to \cite{Ledlie2008}.
The empirical cumulative distribution function of the absolute errors of the prediction, as seen in \cref{fig:vivaldi_abs_error_ecdf}, shows that 
the median absolute error of the predicted latency between clients is 3.37~ms, while 80\% of the experimentation time the nodes had a median error of less than 5~ms.
As far as client-proxy Vivaldi latency prediction is concerned, the median absolute prediction error is 1.07~ms, while 80\% of the experimentation time the nodes had a median error of less than 2.5~ms. \looseness=-1

In our second experiment we tested the ability of the Vivaldi's extended version to adapt to network changes. \Cref{fig:vivaldi_delay_client} shows that there is some delay for Vivaldi to adapt to latency changes between the clients, taking around 30 rounds to adjust its estimates to be over 200~ms. However, as seen in \cref{fig:vivaldi_delay_proxy}, proxy estimates are much faster to adapt, taking around 12 rounds to re-adjust the estimates.\looseness=-1


%


We show that our system can, under real heterogeneous mesh network conditions, estimate the round-trip times between clients as well as between clients and proxies with errors smaller than 5~ms and 2.5~ms respectively. These low prediction and triangulation errors (median relative error in the range of 10\%) are comparable to the original Vivaldi on the Internet. Moreover, we observe that our estimation can eventually trace serious anomalies in the latency of paths. Therefore, we argue that these estimates are satisfactory in order to prioritize paths from clients to proxies that present differences in latency higher than 5~ms and avoid highly loaded paths.\looseness=-1

\textbf{Estimation Based on Other Metrics}
We decided to focus on latency as the most relevant distance metric, considering that the type of web access most essential to users is comprised of typically small HTTP requests and replies (web requests requesting updates to mailboxes, blogs and social network sites, messaging apps, and notifications).
In this context, latency is the most relevant metric to consider in order to assess quality-of-service as perceived by users.\looseness=-1

Considering other web access patterns where larger content dominates (non-essential, though popular), we can adapt the Vivaldi network coordinates system to employ different distance metrics, estimating minimum (or median) sustained throughput (instead of latency) without significant additional overhead. Existing approaches, such as Spruce~\cite{Strauss2003}, exploit the probe gap model (PGM) to collect information about time gaps over consecutive probe packets.
This approach precludes the need for large data transfer to infer throughput that would seriously affect clients, proxies, hindering overall system scalability.
In the specific context of multimedia streaming, related estimation and adaptation techniques for video streaming over HTTP~\cite{Li2014} could be used instead.\looseness=-1

\section{Proxy Performance Estimation}
\label{sec:proxy}


This section shows how TTFB can estimate the current performance of the proxy, expressed as a latency metric, allowing clients to rank choices, avoiding overloaded proxies and proxies with Internet connection that exhibit high delays.\looseness=-1

\textbf{Estimating Proxy Load with TTFB}
TTFB has been widely used in real deployments but also in recent Internet measurement research~\cite{Sundaresan2013,Chen2015} to indicate the responsiveness of a web service since it combines the TCP connection time and the remote server processing time. Moreover, TTFB is a  useful web performance estimator because it is measured passively on the client-side, leveraging information from the already existing client traffic. Nevertheless, our scenario is more complicated, since we aspire to utilize client-side TTFB measurements to estimate the performance of the proxy between the client and the requested content. \looseness=-1

Assuming that $t_{proxy\_ttfb}$ is the time the proxy needs to receive the first byte of response from the server then $t_{response}$ from \cref{eq:4} can also be expressed as \cref{eq:5}. Where $t_{transport\_response}$ is the time until the client has received the complete response. Both $t_{proxy\_ttfb}$ and $t_{transport\_response}$ depend on the available bandwidth of the proxy Internet connection, and the delays in the path from the proxy to the destination server, as well as the responsiveness of the end-server. Additionally, $t_{transport\_response}$ depends on the performance of the client-proxy path. Considering \cref{eq:4}, the TTFB as measured on the client-side can be expressed as \cref{eq:6}.\looseness=-1
\begin{gather}
t_{response\_c\_p} \approx t_{proxy\_ttfb}+t_{transport\_response} \label{eq:5}\\
t_{ttfb\_c\_p} \approx 2*t_{mesh\_rtt\_c\_p} + t_{proxy\_p} + t_{proxy\_ttfb} \label{eq:6}\\
t_{proxy\_p} \approx t_{ttfb\_c\_p} - 2*t_{mesh\_rtt\_c\_p} \label{eq:7} 
\end{gather}

$t_{proxy\_ttfb}$ differs depending on the proxy, the remote server and the requested content. The analysis of the variability of different $t_{proxy\_ttfb}$ latencies,  related to how well the proxies are connected to specific remote servers, lies beyond the scope of this work. Therefore, in our current work we choose not to study $t_{proxy\_ttfb}$ and assume it is stationary for each proxy, representing the delays in the proxy's Internet connection. Nevertheless, as part of our future work we plan to investigate whether and how it is possible to create an estimation model, where each client will be able to use his current and previous HTTP connections to various remote servers in order to identify how this metric affects the measured TTFB. For the rest of this work we assume that all the clients are trying to access the same content that is always available, located in remote servers in similar distance from all the proxies and all the proxies have the same Internet connection bandwidth capacity.\looseness=-1

\begin{figure*}[!t]
\centering
\begin{minipage}[t]{0.30\linewidth}
    \includegraphics[width=1\linewidth]{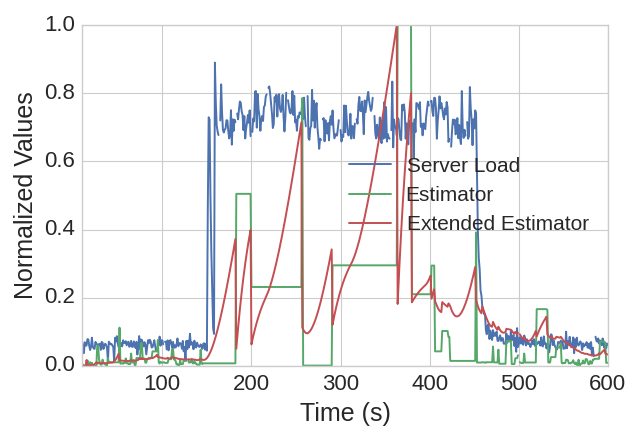}
    \caption{Estimators can assimilate proxy load metrics behaviour ($\alpha = 0.05$) }
    \label{fig:ttfb_time_005_median}
\end{minipage}
\hspace{\columnsep}
\begin{minipage}[t]{0.30\linewidth}
    \includegraphics[width=1\linewidth]{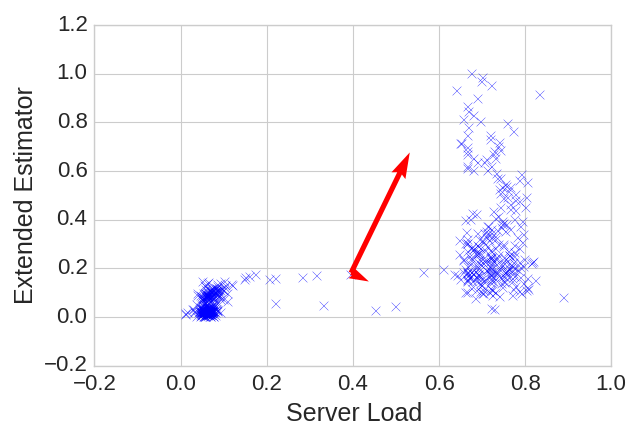}
    \caption{PCA: Extended estimator ($\alpha = 0.05$) can track high proxy load}
    \label{fig:ttfb_pca_005_median}
\end{minipage}
\hspace{\columnsep}
\begin{minipage}[t]{0.30\linewidth}
    \includegraphics[width=1\linewidth]{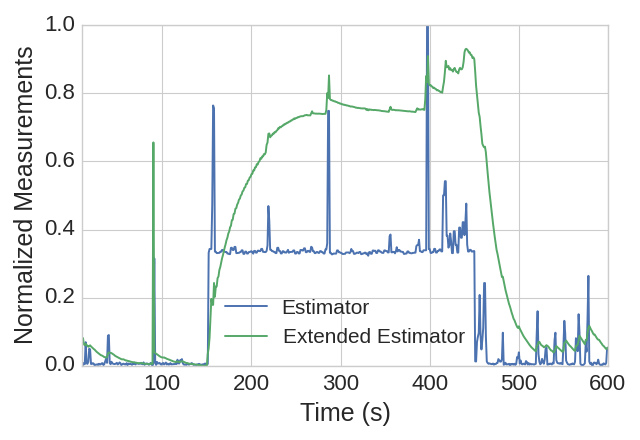}
    \caption{Responsiveness of estimators to Internet connection delays ($\alpha = 0.05$)}
    \label{fig:ttfb_delay_005}
\end{minipage}
\end{figure*}




Therefore, based on \cref{eq:4,eq:6}, the latency incurred by the proxy could be expressed as \cref{eq:7}.
%
%
$t_{proxy\_p}$ can provide an estimation of the proxy performance,  calculated by \cref{eq:7} with the measured TTFB on the client-side and the network. However, the TTFB measurements can be very noisy (sometimes packets are significantly delayed due the proxy or network load, or proxies may complete a request quickly despite heavy load). To minimize the effect of noise on our estimation, we define the \textit{extended TTFB}, where the obtained $t_{proxy\_p}$ values are filtered with an exponential moving average which can be tuned by a parameter $\alpha$. When $\alpha$ is too high, the effect of noise in the measurements leaks into the filtered value, while when $\alpha$ is too low, the filtered values adapt slower to the measured real values, smoothing the peaks and valleys. Moreover, since the TTFB of HTTP requests is measured periodically, we must handle delays that are higher than the measurement period. Therefore, we introduced a penalty scheme to \textit{extended TTFB}, assuming that the request will eventually be completed. Our scheme is based on the simple idea that the TTFB value will be at least as high as the time that the client waited for it. Thus, if a client has not received the first byte for longer than the last $t_{proxy\_p}$ value then the estimated value keeps increasing in every measurement period until it is received.\looseness=-1

Clients periodically exchange the calculated $t_{proxy\_p}$, thus reducing the need for probing, as the value indicates how good a proxy is at serving requests for any client. 
These messages are forwarded through the Vivaldi network. 
Currently, we assume that the client is performing HTTP requests sequentially. 
However, this is not a realistic assumption, since in a typical scenario a browser generates multiple parallel HTTP requests targeting different servers. 
As part of our future work we plan to investigate how to choose or combine measures from multiple HTTP transfers to estimate a TTFB value for each proxy.\looseness=-1 
%

\textbf{Proxy Load Estimation Evaluation}
%
In our first experiment we evaluate the relation between the $t_{proxy\_p}$ and the proxy load. 
The proxy load is represented by various variables monitored on the proxy, including the CPU and the number of incoming and outgoing packets per second in the internal and the external interfaces. \Cref{fig:ttfb_time_005_median} allows the comparison between the normalized median of the proxy variables compared to the estimation and the extended estimation of $t_{proxy\_p}$. 
The proxy is loaded with external requests 150 seconds after the beginning of the experiment and $t_{proxy\_p}$ starts presenting high peaks while the extended estimator presents a more clear relation to the load behavior. 
\Cref{fig:ttfb_pca_005_median} presents another perspective of the relation between the proxy load and the extended estimator, including the plot of the Principal Component Analysis which demonstrates that the higher the load values are the higher the values of the extended estimator.
As a result, we can argue that our extended estimator is behaving similarly to the proxy load, and therefore we can claim it can be used to detect heavily loaded servers.\looseness=-1 

The goal of our second experiment was to evaluate how our estimator responds to proxies having Internet connections with significant delays. Hence, we introduce artificial network delay in the external network interface of the proxy. As seen in \cref{fig:ttfb_delay_005}, both the simple and the extended estimator successfully measure the introduced delay. Nevertheless, the extended estimator appears to need more time to return to the normal levels, as expected. Therefore, we verify that our estimators are responsive to the proxies' Internet connection delays.\looseness=-1

\begin{figure}[!t]
    \centering
    \includegraphics[width=0.6\linewidth]{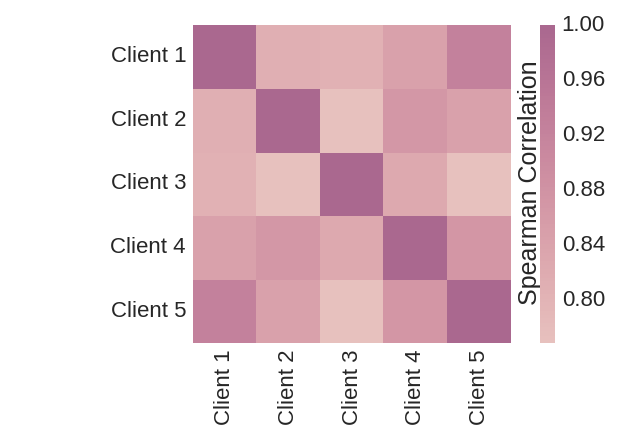}
    \caption{Strong correlation (Spearman's rank) between the clients' Extended TTFB estimators ($\alpha$=0.05)}
    \label{fig:ttfb_heatmap_005}
\end{figure}

\textbf{Sharing Estimations Across Clients}
Despite the fact that our estimators behave similarly with the proxy load, we need to verify that the estimator measured from client $c$ for proxy $p$ can be useful for other clients as well. To investigate this issue, we performed an experiment where one single proxy was used that was serving all the nodes. \Cref{fig:ttfb_heatmap_005} represents the Spearman's rank correlation coefficient~\cite{Spearman1904} between the extended estimators of the different clients throughout the experiment. Spearman's rank correlation coefficient targets to identify correlations that can be expressed by a monotonic function, thus resulting in high values, as we observe in our result, when both of the compared sets ascend or descend similarly.\looseness=-1 



The described proxy performance estimator is not an accurate estimator in terms of absolute values, but has a behavior similar to the proxy load, enabling the client to rank choices and avoid saturated proxies. Moreover, the extended estimator calculated by one client behaves similarly throughout the different clients and can, thus, be disseminated across them reducing the overhead and allowing clients to have updated information concerning proxies they are not currently using.\looseness=-1

\section{Proxy Selection}
\label{sec:evaluation}

After describing our approach for measuring the performance of the network and the proxies, in this section we describe how clients are able to select proxies informed by the presented metrics. Moreover, we present an experiment where clients, adopting our solution, manage to avoid overloaded proxies, very slow internal paths and very slow Internet connections, whereas if a minimum hop or minimum delay selection approach was to be adopted, the clients would not be able to avoid service deterioration.\looseness=-1

On top of our performance estimation tools we built an application-level proxy selection platform. Each client maintains a \textit{proxy selection table}, similarly to a routing table, where each line corresponds to a known proxy and contains the estimated distance, as described in \cref{sec:network}, the extended estimation of the proxy latency, as described in \cref{sec:proxy} and the number of hops to that proxy. Based on this information various proxy selection strategies can be implemented. Nevertheless, the implementations need to take into account the described sensitivity of the provided estimators.\looseness=-1

We used the provided estimators to implement a proxy selection strategy suitable to their rationale, aiming to avoid saturated proxies, proxies with saturated Internet connection as well as proxies behind saturated paths. Therefore, the selection strategy orders the proxies according to the sum of the network latency estimation and the proxy latency estimation, selecting the lowest value. Our implementation avoids unnecessary oscillations by defining a minimum threshold which should be overcome in order to change the selected proxy. Additionally, we implemented a recovery mechanism for situations where a proxy is not being used by any client for a significant amount of time, therefore his current performance estimation value is unknown. In order to prevent all the clients from querying the proxy at the same time, the clients maintain a personalized timeout that depends on a global recovery time, the locally last known measurement of the proxy and their personal network distance to that proxy. If the timeout is reached without receiving any updates, the client is actively probing the proxy to learn its know TTFB value. This way, we manage to make clients that are close to the proxy in charge of querying it and then propagate the information to the other nodes.\looseness=-1

\begin{figure*}[t]
\centering
\begin{minipage}[t]{0.30\linewidth}
    \includegraphics[width=1\linewidth]{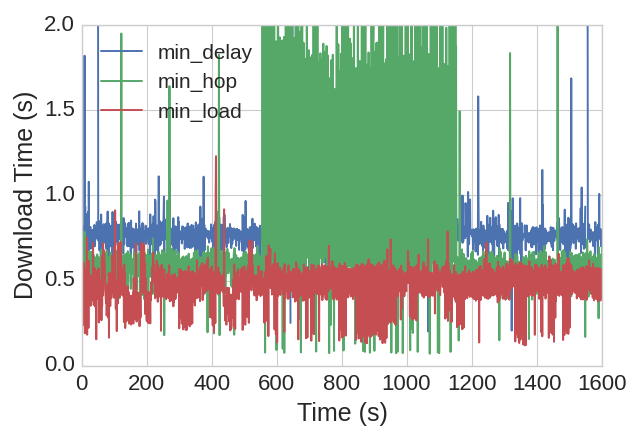}
    \caption{Median client download time for 1Mb per strategy}
    \label{fig:select_rounds}
\end{minipage}
\hspace{\columnsep}
\begin{minipage}[t]{0.30\linewidth}
    \includegraphics[width=1\linewidth]{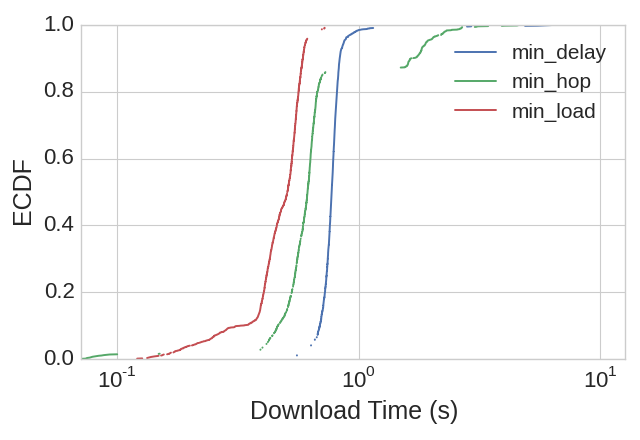}
    \caption{Improvement of median user download time using \texttt{min\_load}  }
    \label{fig:select_ecdf}
\end{minipage}
\hspace{\columnsep}
\begin{minipage}[t]{0.30\linewidth}
    \includegraphics[width=1\linewidth]{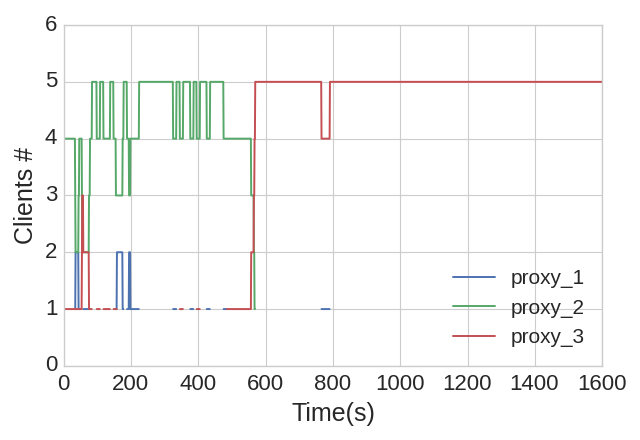}
    \caption{Clients avoid bad choices using \texttt{min\_load} strategy}
    \label{fig:select_mine}
\end{minipage}
\end{figure*}

To evaluate our minimum load selection strategy, following an approach similar to \cite{Ko2013}, we implemented two simple proxy selection strategies based on the minimum hop metric and the minimum network delay metric, that were used to compare to the minimum load solution. Under the minimum hop strategy each client selects the closest proxy in terms of hops while in the minimum network delay strategy the clients select the proxy that has the smallest Vivaldi latency estimator.\looseness=-1

The objective of the evaluation experiment was to describe how the different strategies of the clients deal with the disruptions of the provided service.  The clients use the proxies selected by the routing strategies to repeatedly download files of 1 Mb from the same remote server choosing every 10 seconds, our Vivaldi period, a new proxy if necessary. The value of 1Mb was chosen because in normal conditions a client needs less than the period of 10 seconds to download the file, therefore we can evaluate more accurately the selection alterations. We adopt as evaluation metric the download time experienced by the clients. The experiment lasted 1600 seconds and was repeated for each strategy. Between 50 and 350 seconds we introduce a high amount of requests in one of the proxies. Between 550 and 850 seconds we simulate in one of the proxies an external Internet connection with high delays. Between 1050 and 1350 seconds we simulate a slow network path in one of the proxies.\looseness=-1

\Cref{fig:select_rounds} depicts the median clients' download time per strategy. We observe that our strategy leads the clients to experience a very small amount of download time peaks, especially compared to the static \textit{min\_hops} solution. The y axis of the plot is limited to 2 seconds in order to allow easier comparison, nevertheless the overall distribution of the values can be seen in \cref{fig:select_ecdf}. As depicted, \textit{min\_hop} and \textit{min\_delay} present higher average values compared to \textit{min\_load} (0.76s,0.71s and 0.48s respectively). Most importantly, related to avoiding overloaded options, \textit{min\_hop} and \textit{min\_delay} have many more and significantly higher peaks compared to \textit{min\_load} (maximum 6.33s,4.89s and 1.23 respectively). This is even more apparent for \textit{min\_hop}, that is a static strategy. \textit{min\_load} manages to minimize the number of peaks, confirming our argument that it succeeds to avoid the loaded options. The manner in which \textit{min\_load} is avoiding the loaded options is also shown in \cref{fig:select_mine}, where we observe that clients avoid \textit{proxy\_3} when loaded by requests (150-350 seconds), as well as \textit{proxy\_2} in the ranges of 550-650 seconds and 1050-1350 seconds where we simulate the network path delay and Internet connection delay respectively. It is also worth pointing out that in the performed experiment \textit{min\_delay} and \textit{min\_hop} do not appear to be affected by some of the obstacles introduced, but this is a result of the specific experiment conditions (network latencies and distances) and not of their ability to avoid it.\looseness=-1
%

The results we presented in this section verify how the performance estimators presented in the previous sections can be used by clients to rank and make informed choices from a large set of proxy Internet gateways, avoiding proxies that would deteriorate their user experience.\looseness=-1

\section{Overhead and Scalability Analysis}
\label{sec:overhead}

The two performance estimation components of our system function in parallel. Thus, the total overhead is :\looseness=-1
\begin{align}
\textit{overhead} &= \textit{overhead}_{\textit{vivaldi}} + \textit{overhead}_{\textit{ttfb}}
\end{align}

According to the challenges that Vivaldi~\cite{Dabek2004} faces by design, a network coordinates system should produce a minimal amount of overhead traffic when probing. The overhead network traffic generated by Vivaldi is, in bytes per second:
\begin{gather}
\textit{overhead}_{\textit{vivaldi}} = (2 * \textit{ping}_{\textit{size}} * \textit{ping}_{\textit{freq}}+\textit{data}) * n\\
\textit{data}_\textit{vivaldi} = (n_p*160 + n_n*160 +10)/\textit{round}_{\textit{period}}\\
\textit{ping}_{\textit{freq}} = \textit{round}_{\textit{pings}} / \textit{round}_{\textit{period}}
\end{gather}
In the formulas above, $n$ represents the number of nodes in the Vivaldi system and $n_n$ and $p_n$ are the maximum number of known neighbors and proxies, respectively. We can see that the overhead of Vivaldi increases linearly with the amount of participants.
Vivaldi works in rounds: in every round (which occurs every few seconds) each node sends a few pings to each one of its neighbors. In our deployment we use $8$ pings per round, with a round starting every $10$ seconds. Moreover, in our case it corresponds to one neighbor plus one proxy, and the maximum number of neighbors and proxies is 8. That equates to 436 bytes per second per client, which is acceptable even in a wireless mesh network environment.
For example, assuming all the 30,000 nodes of guifi.net were clients, the overhead would be approximately 1.5 MB/s, distributed all over the network, which sums up to be 1.6\% of the average daily incoming Internet traffic~\cite{Baig2015} (data from 2015).\looseness=-1

The TTFB metrics are passively collected for the proxy currently selected by the client, and then shared between the nodes of the system. Nevertheless, we may ping a proxy if we have not had any metrics for a certain time period as described in \Cref{sec:evaluation}. The network overhead of the proxy TTFB protocol (in bytes per second) is:
\begin{gather}
\textit{overhead}_{\textit{ttfb}} = O(\textit{proxies}) * \textit{payload} / \textit{timeout}\label{eq:overhead_ttfb}\\
\textit{payload} = \textit{payload}_{\textit{request}} + \textit{payload}_{\textit{response}}\\
\textit{timeout} = m_1 * \textit{proxy}_{\textit{distance}} + m_2 * \textit{num\_closer} + b
\end{gather}
Whenever a client overcomes the personalized \textit{timeout} in the proxy information, we query it. 
If we set the $m$ factor too low, the information will not have time to propagate and many nodes will query the proxy. However, if we set the $m$ factor too high, it may take a long time until a node is finally queried.\looseness=-1

Due to the randomly selected neighbors, let us make the assumption that any node may be connected to any other node. Let us assume that a single node pings the proxy. Then, in the next round (assuming a synchronous model), any of the other $N-1$ nodes may query this \textit{knowledgeable} node with probability $1/(N-1)$, pulling the desired proxy information.\looseness=-1

The number of nodes learning the desired information at a given round can be modeled through a binomial distribution with $p=k/N$, where $k$ is the number of nodes that possess such information. The expected value is $k$ -- we expect $k$ nodes learning the information at each round. This means that we expect all nodes, in average, to learn the information after $log_2(N)$ rounds. For the $30,000$ nodes currently registered in guifi.net, that equates to $15$ rounds. 
%
Moreover, it is important to notice that \cref{eq:overhead_ttfb} assumes that \textit{m} and \textit{b} parameters are correctly tuned so that the proxy is contacted by a very low number of nodes with high probability.\looseness=-1




\textbf{Scalability Assessment}
The scalability of our approach stems from four main factors. First, the low client and proxy overhead which was already addressed in this section.
Second, the lack of need for centralized coordination, evident in our approach, having no central coordinator in charge of global decisions. Therefore there are no participants whose processing, state or message load would grow boundless as the number of clients and proxies increase. 
Third factor is bounded storage and traffic, where state size and message count exchanged by each client increase logarithmically with the number of participants. Additionally, the gossip-like propagation of the estimators ensures fast propagation under increasing number of nodes.
The fourth factor is the good convergence of the estimators, where the Vivaldi network estimator is proven to converge 
in large scale scale networks for the selected parameters. 
%
The global convergence of selection is not trivial. We are considering probabilistic strategies from a time perspective as well as individual selection choices that would provide stable aggregate results without further overhead as 
the system scales, given the decentralized nature of our solution that avoids the overhead of global coordination.\looseness=-1


\section{Conclusions}
\label{sec:conclusions}
Communities of citizens develop network infrastructures cooperatively, based on heterogeneous Wireless Mesh Networks. They can achieve global Internet or Web access using a pool of web proxy gateways shared across many participants in the local community network. This affordable Internet access method requires a simple but effective mechanism to arbitrate the client-proxy selection, ensuring a good quality of service and avoiding degradation in the user experience.
This paper introduces reliable and inexpensive latency-based metrics capable of predicting and triangulating performance indicators. It also presents a client-side proxy selection mechanism that combines these metrics to make good choices in terms of QoE or performance, taking into account the contribution of the local network, proxy gateways and their Internet connection. 
This mechanism avoids heavy loaded proxies, proxies with slow Internet connection and slow internal network paths. The overhead is linear to the number of the clients and proxies.\looseness=-1

Future work will explore the sensitivity of proxy delay-based metrics to diverse HTTP traffic. We will utilize multiple HTTP requests to improve proxy performance estimation and use TTFB traffic instead of the additional UDP pings for the client-proxy measures in our extended Vivaldi coordinate system.
We also aspire to extend the system to account for bandwidth-demanding usage, like video streaming and large file downloads, where latency requirements are less relevant.\looseness=-1

\section*{Acknowledgment}

This work was partially supported by the EU funded Erasmus Mundus Joint Doctorate in Distributed Computing (EMJD-DC) (FPA 2012-0030), the EU Horizon 2020 project netCommons (H2020-688768), the Spanish government (TIN2016-77836-C2-2-R), and the Generalitat de Catalunya (2014-SGR-881), the Funda\c{c}\~{a}o para a Ci\^{e}ncia e a Tecnologia (UID/CEC/50021/2013). 

\bibliographystyle{IEEEtran}
\bibliography{IEEEabrv,ms}

\end{document}